\documentclass[sigplan,10pt]{acmart}
\renewcommand\footnotetextcopyrightpermission[1]{}

\usepackage[]{hyperref}

\usepackage{tikz}
\usepackage{float}
\usepackage{amsmath}
\usepackage{xspace}
\usepackage{cleveref}
\usepackage{balance}
\usepackage{url}
\usepackage{siunitx}
\usepackage{comment}
\usepackage{xcolor}
\usepackage{enumitem}
\usepackage{listings}
\usepackage{makecell}
\usepackage{algorithm2e}
\usepackage{multirow}
\usepackage{graphicx}
\usepackage[thicklines]{cancel}
\usepackage{caption}

\author{Yusheng Zheng}
\affiliation{%
  \institution{Eunomia. Inc.}
  \country{USA}
}
\email{yunwei356@gmail.com}
\author{Yiwei Yang}
\affiliation{%
  \institution{UC Santa Cruz}
  \country{USA}
}
\email{yyang363@ucsc.edu}

\author{Haoqin Tu}
\affiliation{%
  \institution{UC Santa Cruz}
  \country{USA}
}
\email{tuisaac163@gmail.com}

\author{Yuxi Huang}
\affiliation{%
  \institution{Eunomia. Inc.}
  \country{USA}
}
\email{yuxi4096@gmail.com}

\title{Code-Survey: An LLM-Driven Methodology for Analyzing Large-Scale Codebases}

\begin{document}




\begin{abstract} 

Modern software systems like the Linux kernel are among the world's largest and most intricate codebases, continually evolving with new features and increasing complexity. Understanding these systems poses significant challenges due to their scale and the unstructured nature of development artifacts such as commits and mailing list discussions. We introduce Code-Survey, the first LLM-driven methodology designed to systematically explore and analyze large-scale codebases. The central principle behind Code-Survey is to treat LLMs as human participants, acknowledging that software development is also a social activity and thereby enabling the application of established social science techniques. By carefully designing surveys, Code-Survey transforms unstructured data—such as commits, emails—into organized, structured, and analyzable datasets. This enables quantitative analysis of complex software evolution and uncovers valuable insights related to design, implementation, maintenance, reliability, and security.

To demonstrate the effectiveness of Code-Survey, we apply it to the Linux kernel's eBPF subsystem. We construct the Linux-bpf dataset, comprising over 670 features and 16,000 commits from the Linux community. Our quantitative analysis uncovers important insights into the evolution of eBPF, such as development patterns, feature interdependencies, and areas requiring attention for reliability and security—insights that have been initially validated by eBPF experts. Furthermore, Code-Survey can be directly applied to other subsystems within Linux and to other large-scale software projects. By providing a versatile tool for systematic analysis, Code-Survey facilitates a deeper understanding of complex software systems, enabling improvements across a variety of domains and supporting a wide range of empirical studies. The code and dataset is open-sourced in https://github.com/eunomia-bpf/code-survey
\end{abstract}

\maketitle

\section{Introduction}

Software systems are increasingly complex, with real-world applications often requiring continuous development over time. Unlike research projects with clearly defined goals and controlled development processes, real-world systems evolve organically. Features are added incrementally, bugs are fixed, and design decisions may be modified years after initial implementation. This complexity is further exacerbated by the need to balance backward compatibility, feature requests, performance improvements, and security patches. As systems evolve, tracing the original intent behind design decisions or understanding the rationale for modifications becomes difficult. This lack of transparency often leads to technical debt, regressions, and challenges in maintaining system reliability.

One of the most prominent examples of a continuously evolving real-world system is the Linux kernel. Serving as the backbone of countless devices and platforms—from cloud servers to mobile devices—the Linux kernel must support a wide array of features while maintaining rigorous performance standards. The \textit{extended Berkeley Packet Filter} (eBPF)\cite{ebpf} subsystem exemplifies this complexity, as it supports critical functionalities such as observability\cite{shen2023network}, networking\cite{vieira2020fast}, and security\cite{deri2019combining}. Despite its significance, much of the development history and design rationale behind eBPF remains underexplored. For example, features like \texttt{bpf\_link}\cite{bpflink}, which provide a new abstraction for attaching programs to events, have been part of the Linux source code for several years but have received little attention outside of kernel developers. Similarly, the increasing use of kfuncs \cite{kfuncs} as replacements for helpers, the growing complexity of control-plane applications, and efforts toward making eBPF Turing complete have not been extensively explored in the broader community.

Understanding the evolution of features in large, complex codebases is a significant challenge in software development\cite{godfrey2008past,mens2008introduction}. Traditional methods, such as static analysis and manual code review, are limited in capturing the full context of a system's growth and change, and require substantial human effort. Unstructured data sources, like commit messages and mailing list discussions, contain valuable insights but are difficult to analyze systematically. Consequently, important information about design decisions, feature evolution, and system behavior is often hidden within large volumes of unstructured text. This makes it nearly impossible to answer questions such as: ``Why was this feature added?'', ``How has this feature evolved?'', or ``What were the discussions that led to this change?''

Recent advancements in artificial intelligence, particularly in Large Language Models (LLMs) like GPT-4o\cite{gpt4o} and O1~\cite{o1}, have opened new opportunities to address these challenges. LLMs have shown great promise in automating software engineering tasks such as code generation\cite{zheng2024kgent}, bug detection\cite{li2024enhancing}, debugging\cite{chen2023teaching}, and error fixing\cite{deligiannis2023fixing}. However, most current applications of LLMs focus on well-defined tasks involving source code or documented APIs. Little work has explored how LLMs can be applied to understand the long-term evolution of large-scale, real-world software systems.

In this paper, we introduce \emph{Code-survey}, a novel methodology that leverages Large Language Models (LLMs) to systematically transform unstructured data—such as commit histories and emails\cite{linux,tan2019communicate,schneider2016differentiating}—into structured datasets for large-scale software analysis. Drawing inspiration from sociological surveys that utilize human participants to gather extensive data, Code-survey employs LLMs to emulate this process, enabling efficient and scalable analysis of software development artifacts. By focusing on the vast amount of text produced during software development, Code-survey allows us to answer questions that were previously difficult to tackle using only structured data in large real-world systems. Through data analysis enabled by Code-survey, we can explore questions such as:

\begin{itemize}
    \item ``How do new feature introductions impact the stability and performance of existing kernel components?''
    \item ``Are there identifiable phases in the lifecycle of a feature, such as initial development, stabilization, and optimization?''
    \item ``How has the functionality of a specific eBPF feature evolved over successive commits?''
    \item ``Which components or files in the Linux kernel have the highest bug frequency?''
    \item ``What lessons can be learned from the development history of kernel eBPF that can be applied to improving other eBPF runtimes?''
    \item ``What dependencies have emerged between features and component?''
\end{itemize}

To demonstrate the efficacy of Code-survey, we apply it to the \textit{Linux-bpf dataset}, which contains over 670 features, 15,000 commits, and 150,000 emails related to the development of the eBPF subsystem. Through this structured dataset, we uncover new insights into the design and evolution of features like \texttt{bpf\_link}, and highlight trends that were previously hidden within the unstructured data. These insights have been initially confirmed by eBPF experts.

The key contributions of this paper are as follows:

\begin{itemize}
    \item We introduce \emph{Code-survey}, a novel methodology that leverages LLMs to transform unstructured data produced in software development into structured datasets via surveys, enabling systematic analysis of software evolution. To the best of our knowledge, \emph{Code-survey} is the first methodology that leverages LLMs for the systematic analysis of large-scale codebases.
    \item We present the \textit{Linux-bpf dataset}, a structured dataset comprising over 670 features, 15,000 commits, and 150,000 emails related to the eBPF subsystem in the Linux kernel.
    \item We apply the \emph{Code-survey} methodology to build an LLM-driven agent system, allowing us to perform systematic analysis on the \textit{Linux-bpf dataset}.
    \item We demonstrate that the \emph{Code-survey} methodology reveals new insights into the evolution of eBPF kernel features that are impossible to uncover using traditional methods. By combining traditional data analysis methods with eBPF experts' domain knowledge and historical context, we also initially verified the consistency and correctness of the data.
    \item We identify and highlight under-explored areas of eBPF to support various use cases with the help of \emph{Code-survey}, pointing out interesting research directions.
\end{itemize}

The remainder of this paper is structured as follows. We review background in Section~\ref{sec:related}, followed by a detailed explanation of the Code-survey methodology in Section~\ref{sec:methodology}. Section~\ref{sec:analysis} presents our analysis of the Linux-bpf dataset and the insights gained from it. Section~\ref{sec:best_practices} discusses best practices when designing AI-based surveys. Sections~\ref{sec:limitations} and~\ref{sec:future} conclude with a discussion of current limitations and future work. All artifacts are open-sourced at \url{https://github.com/eunomia-bpf/code-survey}.

\section{Background}
\label{sec:related}

This section discuss the role of Large Language Models in software development, the complexities of Linux kernel development, and the importance of survey methodologies in empirical software engineering research.

\subsection{LLMs in Software Development}

Large Language Models (LLMs), such as GPT-4o and Claude, have significantly impacted software development~\cite{jin2024llms}, particularly in automating tasks like code generation, debugging, and testing. Tools like GitHub Copilot~\cite{copilot} leverage these models to enhance developer productivity by providing intelligent code suggestions. Despite these advancements, challenges such as hallucinations—where incorrect but plausible code is generated—still persist~\cite{fan2023large,ji2023survey}. Moreover, most research focuses on well-defined tasks, and exploring the design and evolution of large-scale software systems using LLMs remains underexplored.

Additionally, using LLMs for summarizing test results, decision-making, and converting unstructured data into structured formats is becoming increasingly common in both academia~\cite{jin2024comprehensive,iourovitski2024grade,patel2024lotus} and industry~\cite{llmnvida}. This capability is especially valuable in environments where massive amounts of unstructured data—such as logs, emails, or messages—exist. The ability to systematically extract insights from this data enables more efficient analysis and has been widely applied to tasks like market research~\cite{brand2023using}.

\subsection{Software Evolution and Its Challenges}
Software evolution involves the continuous modification and adaptation of systems to meet changing requirements~\cite{lehman1996laws}. According to Lehman's laws, systems must evolve to remain useful, but this often increases complexity without proactive management. In large-scale systems like the Linux kernel~\cite{linux}, evolution is non-linear, involving numerous contributors and revisions, which leads to intricate interdependencies~\cite{israeli2010linux}. The Linux kernel generates vast unstructured data, including commit logs and email threads, which traditional analysis methods struggle to process~\cite{mens2008introduction}. These artifacts contain rich contextual information about design decisions, but their volume and unstructured format hinder conventional techniques. Additionally, evolving systems accumulate \emph{technical debt}~\cite{brown2010managing}, increasing maintenance costs and reducing reliability. Addressing these challenges requires innovative approaches capable of handling large-scale unstructured data to provide actionable insights.

\subsection{Survey Methodology and Empirical Studies}

Empirical studies in software engineering~\cite{perry2000empirical} are crucial for understanding how software evolves and how development practices affect system reliability, performance, and maintainability. Traditional surveys rely on structured questionnaires or interviews but are limited by scale and biases such as subjective recall or incomplete responses. In contrast, the \textit{Code-Survey} methodology automates data collection using LLMs to extract structured insights from unstructured data like commit histories and mailing lists. This approach enables large-scale analysis and allows us to answer questions that traditional methods cannot.

\section{\emph{Code-Survey} Methodology}
\label{sec:methodology}

This section outlines the \emph{Code-Survey} methodology, which integrates the expertise of human participants and the efficiency of Large Language Model (LLM) agents to convert unstructured software data—such as commit histories and mailing lists—into structured datasets suitable for analysis.

The central principle behind \emph{Code-Survey} is to treat LLMs as human participants in a survey-like process. While LLMs can process data more quickly and at lower costs than humans, they are also prone to errors, guesses, and limitations in specific domains, much like human participants~\cite{ji2023survey}. By leveraging traditional survey methodologies designed for humans, we can efficiently conduct LLM-based surveys while maintaining oversight and validation from human experts. Additionally, LLM agents can utilize traditional tools to automate analysis tasks.

\begin{figure}[t]
    \centering
    \includegraphics[width=0.5\textwidth]{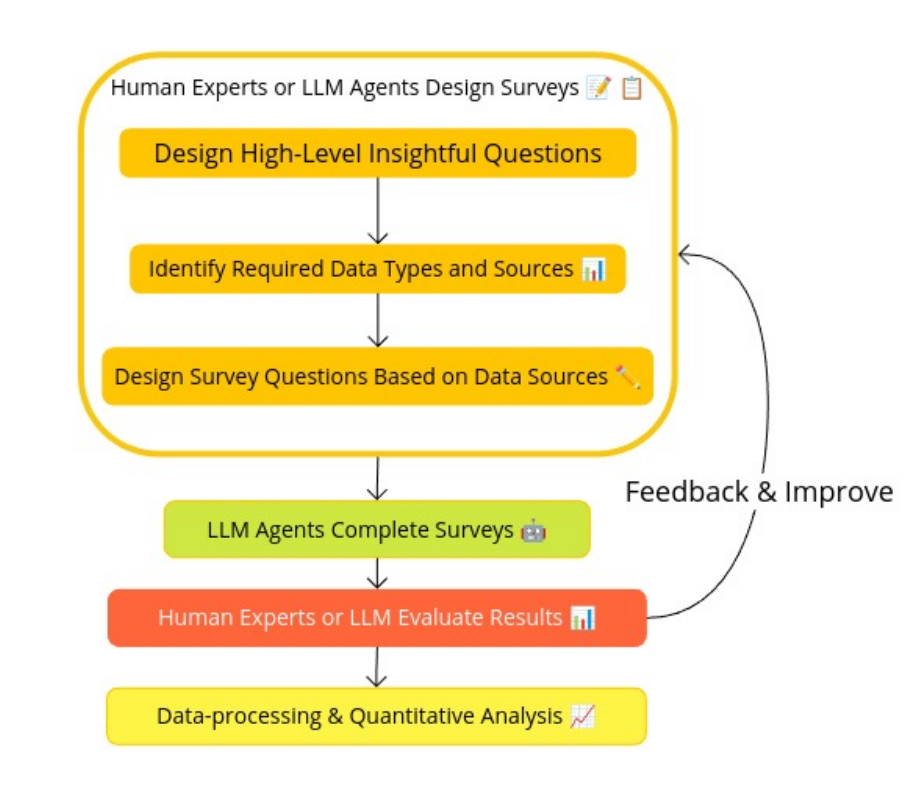}
    \caption{\emph{Code-Survey} Workflow}
    \label{fig:workflow}
\end{figure}

\subsection{Workflow}

The \emph{Code-Survey} process follows a structured workflow, as illustrated in Figure~\ref{fig:workflow}:

\begin{enumerate}
    \item \textbf{Survey Design by Human Experts or LLM Agents}: Experts or LLM agents create tailored questions for each type of unstructured data (e.g., commits, emails) to extract key insights. The survey design is crucial for guiding the LLM in structuring the data effectively and focuses on tasks aligned with the LLM's strengths, such as summarization and yes/no questions. This step should minimizes the usage of open-ended questions that require deep domain expertise.

    \item \textbf{Survey Completion by LLM Agents}: LLM agents process the unstructured input by answering the survey questions. They organize data into structured formats by tagging relevant information, summarizing key points, and identifying important patterns. This step is fundamental in transforming unstructured data into structured datasets, enabling more straightforward analysis.

    \item \textbf{Evaluation of Survey Results by Human Experts or LLM Agents}: Human experts with domain-specific knowledge, or additional LLM agents, evaluate a sample of the structured data. This evaluation ensures accuracy and allows for the detection of discrepancies. If the results are unsatisfactory, the process loops back to Step 1, where the survey design is refined.

    \item \textbf{Data Processing and Quantitative Analysis}: The structured data undergoes further processing to analyze key metrics and patterns. Quantitative analysis is performed to identify trends in development, feature interdependencies, and areas needing improvement in reliability and security. This provides a detailed view of the system's evolution and characteristics.

\end{enumerate}

This process employs an iterative design, allowing surveys to be refined based on analysis results, thereby creating a feedback loop that enhances data structuring and interpretation. By leveraging both LLMs and human oversight, the \emph{Code-Survey} methodology efficiently handles large volumes of unstructured software data. At each step, human experts can interactively contribute, or LLM agents can automate the process.

\subsection{Survey Design with LLM Assistance}

A key aspect of \emph{Code-Survey} is designing effective surveys to generate accurate data. Surveys can be designed by humans or LLM agents. We identify three key steps to guide LLM agents in designing surveys. The following prompts serve as a framework or LLM input for survey creation:

\begin{enumerate}
    \item \textbf{Design High-Level Insightful Questions}: If you could ask every kernel developer to complete a survey or questionnaire about a commit or an email, what are the most insightful questions related to design, implementation, maintenance, reliability, and security? Describe the possible questions in detail.

    \item \textbf{Identify Required Data Types and Sources}: What data types and sources are required to answer the insightful questions described previously? Describe the data types and sources for each question in detail.

    \item \textbf{Design Survey Questions to Retrieve Data}: What survey questions can you design to obtain the required data types for the insightful questions from the data sources described previously? Describe the survey questions for kernel developers in detail.

\end{enumerate}

This workflow ensures that the LLM-driven survey design process leads to structured data that offers deeper insights into complex software systems, such as the Linux kernel. By guiding LLM agents through these steps, we can systematically extract valuable information from unstructured data sources.

\section{Case Study: eBPF}
\label{sec:analysis}

In this section, we apply the \emph{Code-Survey} methodology to analyze the evolution and development of the eBPF subsystem in the Linux kernel. Our goal is to uncover insights into the lifecycle, stability, and design decisions of key eBPF features. The survey results are validated through expert review and both quantitative and qualitative analyses.

The Extended Berkeley Packet Filter (eBPF)~\cite{ebpf} is a rapidly evolving subsystem in the Linux kernel that allows users to run sandboxed programs in kernel space without modifying the kernel itself~\cite{lim2024safebpf}. Originally developed for packet filtering, eBPF now supports diverse use cases such as performance tracing, security monitoring, and system observability, and has been expanded to multiple platforms~\cite{windows-ebpf,zheng2023bpftime}. The academic community has identified current problems and limitations of eBPF, proposing several works to improve aspects like the verifier and deployment.

\subsection{Motivation: Analyzing eBPF Kernel Features with Data Analysis}

To understand the limitations of traditional data analysis methods, we begin by analyzing the evolution of eBPF kernel features using well-defined kernel commits. Traditional methods allow us to study the development of various eBPF features, such as helpers, maps, attach types, and other functionalities. We obtained feature pairs from eBPF documentation~\cite{ebpfdocs} and combined them with Git commit data.

\subsubsection{How Do All eBPF Features Evolve Over Time?}

\begin{figure}[ht]
    \centering
    \includegraphics[width=\linewidth]{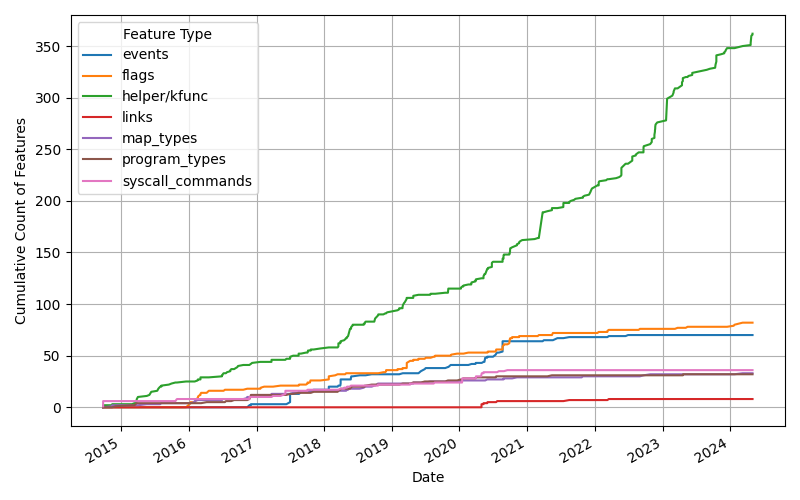}
    \caption{Cumulative eBPF Features Timeline}
    \label{fig:cumulative_feature_timeline}
\end{figure}

Figure~\ref{fig:cumulative_feature_timeline} shows trends in the adoption of new features within the overall eBPF subsystem. Helpers and kfuncs exhibit the most significant growth, supporting various use cases and changes, while other features have shown steady increases over the past four years, indicating their maturation.

\subsubsection{Timeline of eBPF Helper Functions vs.\ Kfuncs}

\begin{figure}[ht]
    \centering
    \includegraphics[width=\linewidth]{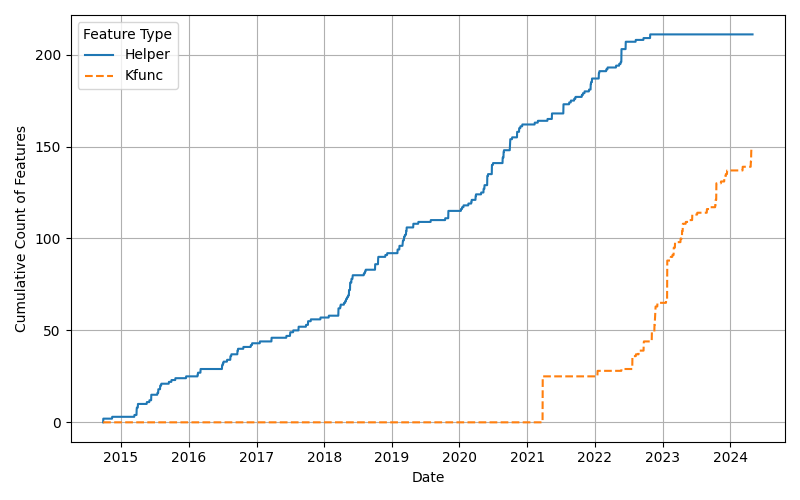}
    \caption{Cumulative Helper and kfunc Timeline}
    \label{fig:cumulative_helper_kfunc_timeline}
\end{figure}

Figure~\ref{fig:cumulative_helper_kfunc_timeline} illustrates the evolution of eBPF helper functions and kfuncs over time. Since 2023, helper functions have remained stable with almost no new additions, whereas kfuncs are growing rapidly; this demonstrates the community's interest in expanding kernel interaction via kfuncs, and all new use cases now tend to use kfuncs to influence kernel behavior, signaling a shift toward deeper kernel integrations.

\subsubsection{What Are the Patterns in Other eBPF Features?}

\begin{figure}[ht]
    \centering
    \includegraphics[width=\linewidth]{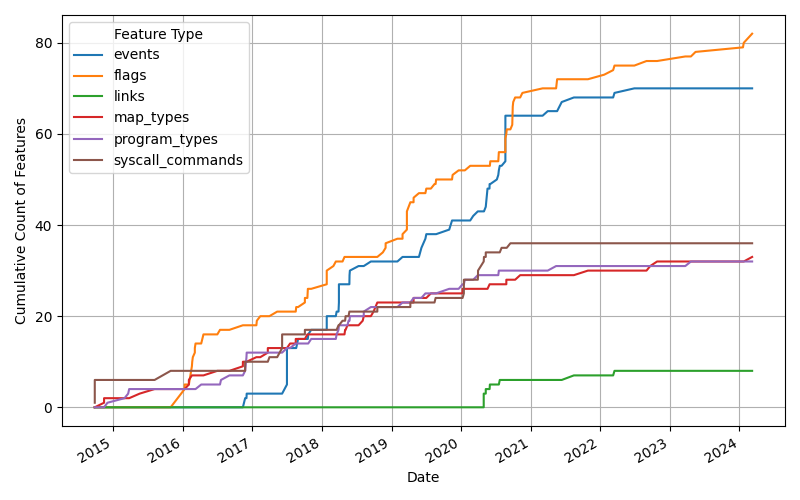}
    \caption{Cumulative eBPF Features Timeline Without Helpers and kfuncs}
    \label{fig:cumulative_without_helper_timeline}
\end{figure}

Figure~\ref{fig:cumulative_without_helper_timeline} examines the evolution of eBPF features, excluding helpers and kfuncs, with a focus on core eBPF functionalities and their impact on the overall subsystem. After 2020, core features such as events, flags, map types, and program types have stabilized. Notably, the introduction of \textbf{bpf\_link} coincides with the effort on the management of growing use cases and complexity, resulting in a significant difference observed before and after its introduction.

\paragraph{However\ldots}

The analysis of eBPF features mainly relies on kernel structured definitions and data sources obtained by humans, which is limited by data availability and the time-consuming process of manual data collection.

There is a massive amount of unstructured data in the form of commit messages, code changes, and developer discussions that can provide valuable insights into the evolution and design decisions of eBPF features. While it is possible for humans to perform empirical analysis on some of this data, covering all of it is practically \emph{impossible}.

\textbf{Can AI assist us?} Instead of relying on large language models (LLMs) to attempt kernel coding—which may yield incorrect answers—we propose a different quantitative approach.

By carefully designing a \emph{survey} and utilizing LLMs to \emph{transform} unstructured data like commits and emails into well-organized, structured, and easy-to-analyze datasets, we can perform \textbf{quantitative} analysis using traditional methods. In this way, AI can help analyze data and provide answers quickly; this capability is already a feature of platforms like ChatGPT.

\subsection{Research Questions for Survey Evaluation}

To ensure that the LLM does not answer questions randomly, we evaluate the effectiveness and correctness of the \emph{Code-Survey} results by exploring the following high-level research questions:

\begin{itemize}
    \item \textbf{Correctness of Survey Responses:} How can we ensure that survey responses reflect accurate and relevant information about the system's features and commits?
    \item \textbf{Consistency Across Similar Questions:} Are similar questions answered consistently across different but related features or subsystems?
    \item \textbf{Coverage of Survey Questions:} Do the survey questions comprehensively cover all relevant aspects of the feature or subsystem under analysis?
    \item \textbf{Insight from Survey:} Can the survey data help users analyze the design, implementation, maintenance, reliability, and security evolution, and gain valuable insights?
    \item \textbf{Clarity and Ambiguity in Responses:} Are the survey responses clear and unambiguous, making them actionable for further analysis?
    \item \textbf{Alignment with Real-World Changes:} How accurately do survey results reflect real-world changes in the software's features and evolution?
    \item \textbf{Expert Confirmation:} How do experts rate the accuracy of the survey's generated insights?
\end{itemize}

\subsection{Commit Survey Design and Objectives}

To gain deeper insights into the design and evolution of the eBPF subsystem, we developed a comprehensive survey aimed at classifying commits within the Linux eBPF subsystem. This survey evaluates specific aspects of each commit by analyzing commit messages and associated code changes. In cases where the commit message provides insufficient information, respondents are allowed to select an ``I'm not sure'' option.

The primary goals of the survey are to:

\begin{itemize}
    \item Provide concise summaries of each commit.
    \item Extract key themes and components affected by each commit.
    \item Classify commits based on their type, complexity, and impacted components.
    \item Identify patterns and trends in the evolution of the eBPF subsystem.
\end{itemize}

\subsubsection{Survey Structure}

The survey consists of a series of structured questions designed to capture the essential characteristics of each commit. Respondents are encouraged to be as specific as possible based on the available commit message and code changes. If the commit message lacks clarity, the ``I'm not sure'' option can be selected.

\subsubsection{Simplified Survey Definition}

Below is the simplified survey structure:

\begin{enumerate}
    \item \textbf{Summary}: Provide a one-sentence summary of the commit (max 30 words).
    \item \textbf{Keywords}: Extract up to three keywords from the commit.
    \item \textbf{Commit Classification} (\emph{Single Choice}): What is the main type of the commit?
    \begin{enumerate}[label=(\alph*)]
        \item Bug fix
        \item New feature
        \item Performance optimization
        \item Code cleanup or refactoring
        \item Documentation change or typo fix
        \item Test case or test infrastructure change
        \item Build system or CI/CD change
        \item Security fix
        \item Merge commit
        \item Other type of commit
        \item I'm not sure
    \end{enumerate}
    \item \textbf{Commit Complexity} (\emph{Single Choice}): Estimate the complexity of implementing this commit.
    \begin{enumerate}[label=(\alph*)]
        \item Simple (affects 1--20 lines or 1--2 files)
        \item Moderate (affects 21--100 lines or a few files)
        \item Complex (affects over 100 lines or 5+ files)
        \item Merge-like (merges multiple branches or features)
        \item Non-code or generated changes
        \item I'm not sure
    \end{enumerate}
    \item \textbf{Major Related Implementation Component} (\emph{Single Choice}): What is the main implementation component modified?
    \begin{enumerate}[label=(\alph*)]
        \item eBPF verifier
        \item eBPF JIT compiler
        \item Helpers and kfuncs
        \item Syscall interface
        \item eBPF maps
        \item \texttt{libbpf} library
        \item \texttt{bpftool} utility
        \item Test cases and makefiles
        \item Changes in other subsystems related to eBPF events
        \item Merge commit
        \item Other component related to eBPF
        \item Unrelated to eBPF subsystem
        \item I'm not sure
    \end{enumerate}
    \item \textbf{Major Related Logic Component} (\emph{Single Choice}): What is the main logic component affected?
    \begin{enumerate}[label=(\alph*)]
        \item eBPF instruction logic
        \item Runtime features logic
        \item eBPF events-related logic
        \item Control plane interface logic
        \item Maps logic
        \item BPF Type Format (BTF) logic
        \item Merge commit
        \item General utilities logic
        \item Other eBPF logic component
        \item Unrelated to eBPF subsystem
        \item I'm not sure
    \end{enumerate}
    \item \textbf{Use Cases or Submodule Events} (\emph{Multiple Choice}): What eBPF use cases or subsystem events does the commit relate to?
    \begin{enumerate}[label=(\alph*)]
        \item XDP-related programs
        \item Socket-related programs
        \item \texttt{tc}-related programs
        \item Netfilter-related programs
        \item Tracepoints-related programs
        \item Kernel probes (\texttt{kprobe}/\texttt{ftrace})
        \item User-space probes (\texttt{uprobe}/USDT)
        \item Profiling-related programs
        \item LSM-related programs
        \item \texttt{struct\_ops}-related programs
        \item \texttt{cgroup}-related programs
        \item HID driver-related programs
        \item Scheduler-related programs
        \item Improves overall eBPF infrastructure
        \item Merge commit
        \item Other eBPF use cases
        \item Unrelated to eBPF subsystem
        \item I'm not sure
    \end{enumerate}
\end{enumerate}

\subsection{Implementing the Survey Using LLM Agents}

To efficiently process and analyze the vast number of commits in the Linux eBPF subsystem, we leveraged LLMs to automate our survey. We developed Assistant Agents using GPTs~\cite{gpts} for generating survey responses and assisting in analyzing results. By utilizing the GPT-4o~\cite{gpt4o} LLM model, we transformed unstructured commit data into structured responses aligned with our survey definitions. We initially applied this method to over 15,000 commits spanning eight years and plan to expand it to include emails and patches.

\subsubsection{Commit Survey Methodology}

Our approach uses an AI model to analyze each commit's details and answer the survey questions. The process includes:

\begin{enumerate} 
\item \textbf{Data Extraction}: We gather key commit details such as commit ID, author information, commit date, message, code changes, and associated emails.
\item \textbf{Prompt Construction}: A prompt containing the survey title, description, changed files, and commit details is generated to guide the AI.
\item \textbf{AI Model Interaction}: The AI processes the prompt, analyzing the commit message and code changes to respond to the survey. For each commit, the AI receives the commit details and survey questions, completes them, and generates structured output as JSON.
\item \textbf{Feedback Loop}: If the AI's response is incomplete or inconsistent, it re-evaluates and revises the answers to improve accuracy.
\item \textbf{Data Aggregation}: The AI's responses are stored for later quantitative analysis.
\end{enumerate}

We used the GPT-4o model for its strong language understanding and ability to handle technical content, making it well-suited for analyzing kernel commits.

\subsubsection{Enhancing Survey Accuracy}

Due to time and budget limitations, the AI's performance in this experiment can still be greatly improved. Several strategies can enhance accuracy:

\begin{itemize}
\item \textbf{Multiple Iterations}: Running the survey multiple times and averaging results, similar to human surveys, can significantly reduce response variability. Due to budget limits, we ran it once per commit, and the results are already meaningful with less than 1\% error. We plan to run the survey multiple times and conduct a better evaluation in future work.
\item \textbf{Advanced Models}: Fine-tuning domain-specific LLMs for kernel development, or using more advanced models like O1~\cite{o1}, can capture technical nuances better.
\item \textbf{Refined Prompt Engineering}: Crafting clearer prompts and survey questions will improve response accuracy.
\item \textbf{Multi-Step Reasoning}: Using multi-step processes or multi-agent systems can help analyze complex commits more thoroughly.
\end{itemize}

Automating the survey with LLMs enabled us to efficiently process tens of thousands of commits, transforming unstructured data into structured insights.

\subsection{The Commit Dataset}

The \texttt{commit\_survey.csv} dataset provides detailed metadata for over 15,000 Linux kernel commits, including commit types, messages, timestamps, and affected components. It is used to categorize and classify commits, focusing on the eBPF subsystem.

\subsubsection{Dataset Overview}

The dataset contains the following fields:

\begin{itemize}
    \item \textbf{Commit Metadata}: Unique commit IDs, author and committer details, and timestamps.
    \item \textbf{Commit Messages and File Changes}: Descriptions of the changes in each commit.
    \item \textbf{Classification}: Types such as bug fixes, feature additions, or merges.
    \item \textbf{Complexity}: Based on the number of files and lines changed.
    \item \textbf{Components}: Affected implementation and logic components.
    \item \textbf{Use Cases}: Related subsystems and modules.
\end{itemize}


\paragraph{Commit Classification Distribution}

\begin{figure}[ht]
    \centering
    \includegraphics[width=\linewidth]{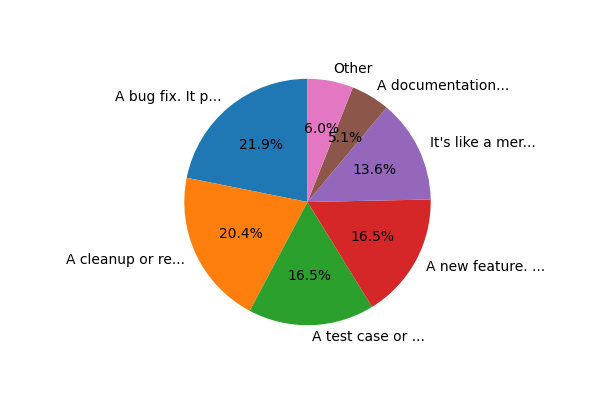}
    \caption{Commit Classification Distribution}
    \label{fig:commit_pie_chart_commit_classification}
\end{figure}

Most commits focus on bug fixes and code cleanups, reflecting ongoing efforts to maintain code quality. Significant attention is also given to testing infrastructure changes, emphasizing the importance of robustness. New features constitute a considerable portion of the commits, while merge commits are commonplace in the Linux kernel's development process.

\paragraph{Commit Complexity Distribution}

\begin{figure}[ht]
    \centering
    \includegraphics[width=\linewidth]{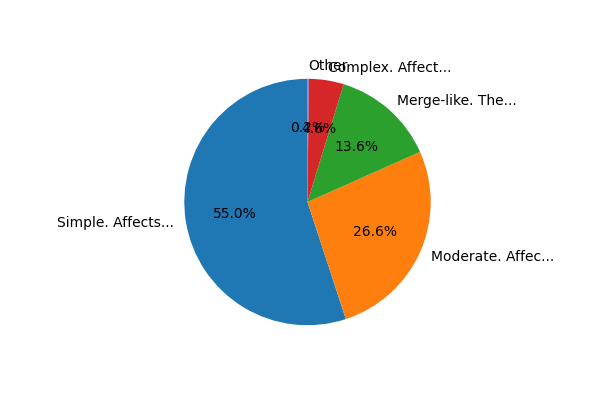}
    \caption{Commit Complexity Distribution}
    \label{fig:commit_pie_chart_commit_complexity}
\end{figure}

The majority of commits are simple, involving small changes, while more complex changes constitute a smaller but noteworthy portion.

\paragraph{Major Related Implementation Components}

\begin{figure}[ht]
    \centering
    \includegraphics[width=\linewidth]{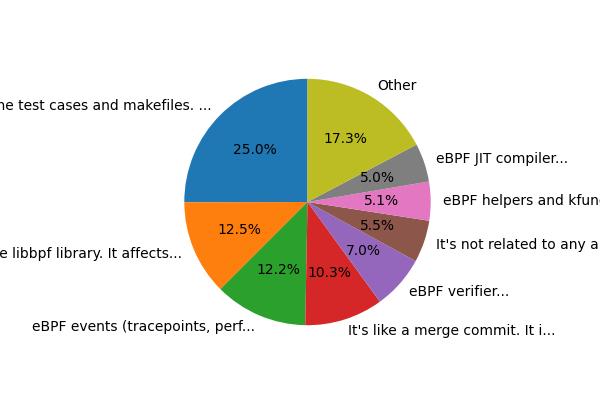}
    \caption{Major Related Implementation Components}
    \label{fig:commit_pie_chart_major_implementation_component}
\end{figure}

Test cases and build scripts are significantly affected, highlighting continuous improvements in testing and build processes. The \texttt{libbpf} library is also a key component in the kernel eBPF toolchain, receiving considerable development attention. Additionally, substantial development occurs in other kernel subsystems, particularly related to eBPF events. The frequent updates to the eBPF verifier and helpers indicate ongoing efforts to enhance functionality and ensure program safety. Notably, some commits appear unrelated to the eBPF subsystem.

\paragraph{Logic Components Affected}

\begin{figure}[ht]
    \centering
    \includegraphics[width=\linewidth]{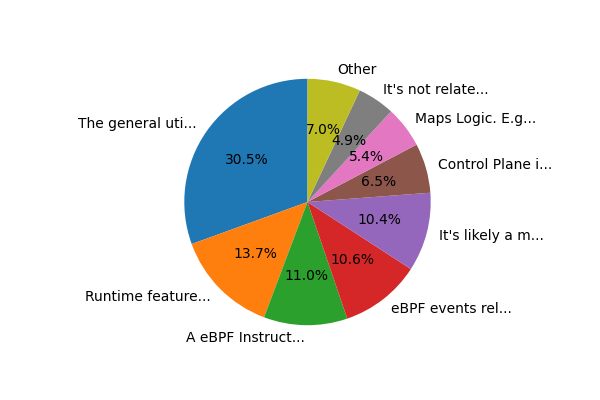}
    \caption{Major Related Logic Components}
    \label{fig:commit_pie_chart_major_logic_component}
\end{figure}

General utilities, such as tools and scripts, receive the most updates, followed by runtime features like helpers and kernel functions, which are consistently enhanced. eBPF event logic and instruction handling are also frequently updated to ensure robustness and functionality.

\paragraph{Use Cases and eBPF Events}

\begin{figure}[ht]
    \centering
    \includegraphics[width=\linewidth]{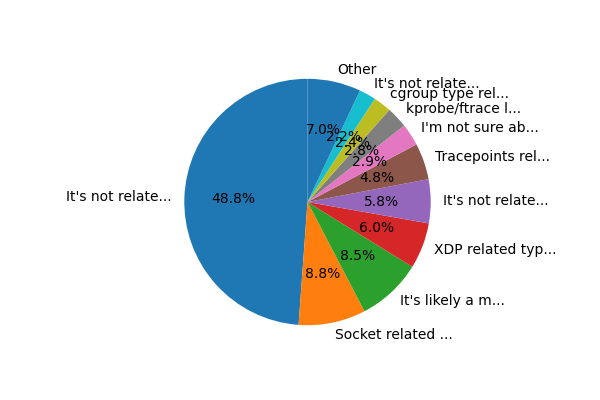}
    \caption{Related Events and Use Cases}
    \label{fig:commit_pie_chart_usecases_or_submodule_events}
\end{figure}

While most commits enhance the core eBPF infrastructure—including the verifier and runtime components—significant development also extends to networking-related features such as socket and XDP programs, which receive substantial attention. Additionally, tracing tools like tracepoints and kprobes highlight eBPF's crucial role in system diagnostics and debugging.

\subsection{Correctness of Survey Responses}

To ensure that the survey responses accurately reflect system features and commits, we validated the results by randomly sampling the data and cross-referencing with expert knowledge and processing logs. Specifically, we addressed two common issues: the handling of merge commits and the identification of commits unrelated to the eBPF subsystem.

\subsubsection{Merge Commit Correctness}

We observed discrepancies in how merge commits were classified between the commit classification and the major implementation component perspectives. To address this, we analyzed the top merge commit messages and found that some merge commits were categorized based on their predominant effect on a specific component rather than being uniformly labeled as merge commits. By clarifying this distinction in the survey questions, we ensured that the classification system accurately reflected the commit's impact across different perspectives.

\textbf{Example:}
\begin{verbatim}
Top Commit (Classification but not Implementation):
3    Merge branch 'bpf-fix-incorrect-name-check-pass'
25   Merge branch 'vsc73xx-fix-mdio-and-phy' Pawel...
\end{verbatim}

\subsubsection{Non-Related eBPF Subsystem Commits}

We noted that commits not directly related to the eBPF subsystem were sometimes included due to broad filtering criteria (e.g., \texttt{--grep=bpf}). Upon reviewing these commits, we confirmed that some mentioned ``bpf'' but addressed unrelated or peripheral issues.

\textbf{Example:}
\begin{verbatim}
Sample 'Not related to eBPF' Commit Messages:
17    bonding: fix xfrm real_dev null pointer...
21    btrfs: fix invalid mapping of extent xarray st...
\end{verbatim}

\subsubsection{Consistency Across Similar Questions}

We checked for consistency in responses by comparing related questions, focusing on the number of merge commits in commit classifications and complexities, as well as the number of commits unrelated to the eBPF subsystem in the implementation and logic components.

\textbf{Example 1: Merge Commits}
\begin{itemize}
    \item \textbf{Commit Classification:} ``It's like a merge commit.'' (2,130 responses)
    \item \textbf{Commit Complexity:} ``Merge-like. The commit merges branches.'' (2,132 responses)
\end{itemize}

\textbf{Example 2: Unrelated Components}
\begin{itemize}
    \item \textbf{Implementation Component:} ``Not related to eBPF.'' (773 responses)
    \item \textbf{Logic Component:} ``Not related to eBPF.'' (766 responses)
\end{itemize}

The close alignment of these numbers demonstrates consistent identification of unrelated components. With a low misclassification rate (under 0.05\% for total commits), our data shows high consistency, supporting the reliability of the survey design.

\subsection{Timeline Analysis of Commits}

Analyzing the timeline of commits provides valuable insights into the evolution of the eBPF subsystem over time. By visualizing the distribution of commit types, complexities, and major components across different periods, we can identify trends and patterns in the development of eBPF features.

The data was processed by cleaning to remove irrelevant commits, smoothing using a 3-month average to reduce noise and highlight long-term trends, and treating single-component \texttt{Merge} commits as regular commits while removing multi-component \texttt{Merge} commits, such as mainline merges. The time span covers from 2017 to the end of 2024, encompassing over 15,000 commits.

\subsubsection{Commit Classification Over Time}

\begin{figure}[ht]
    \centering
    \includegraphics[width=\linewidth]{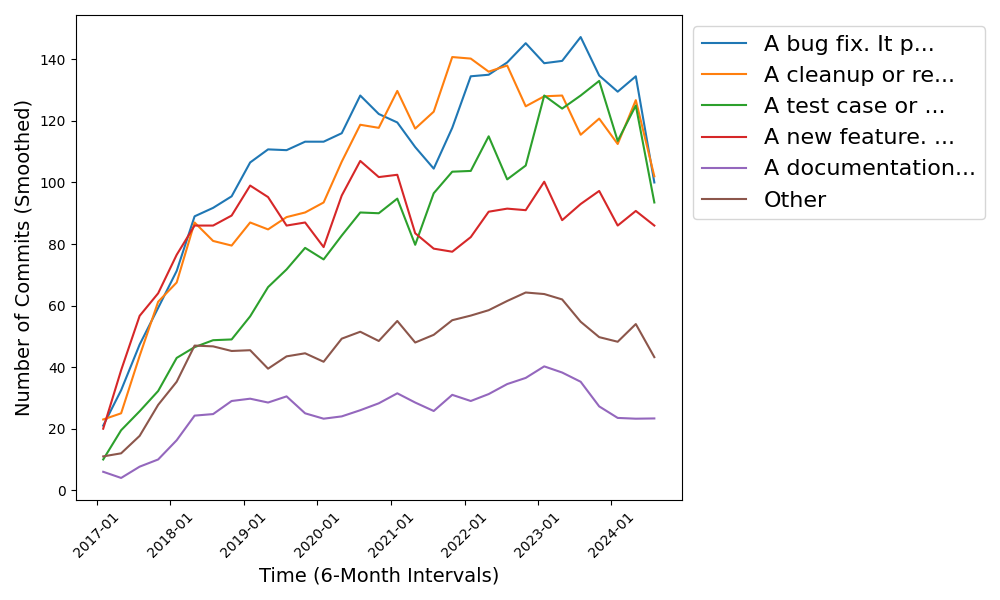}
    \caption{Commit Classification Over Time}
    \label{fig:timeline_commit_classification_smoothed}
\end{figure}

Figure~\ref{fig:timeline_commit_classification_smoothed} shows the distribution of different types of commits over time.

The development began to grow significantly in 2017, with limited addition of test cases initially, which continued to improve over time. New feature development follows a cyclical pattern, with notable spikes around 2020 and 2021. After 2021, the number of cleanups and refactorings increases significantly, while new feature additions decline, indicating a shift in focus towards code maintainability and stability. A decline in cleanup commits after 2023 suggests that while new features continue to be added, the emphasis has shifted more towards stabilization and optimization.

\subsubsection{Commits Related to Major Implementation Components Over Time}

\begin{figure}[ht]
    \centering
    \includegraphics[width=\linewidth]{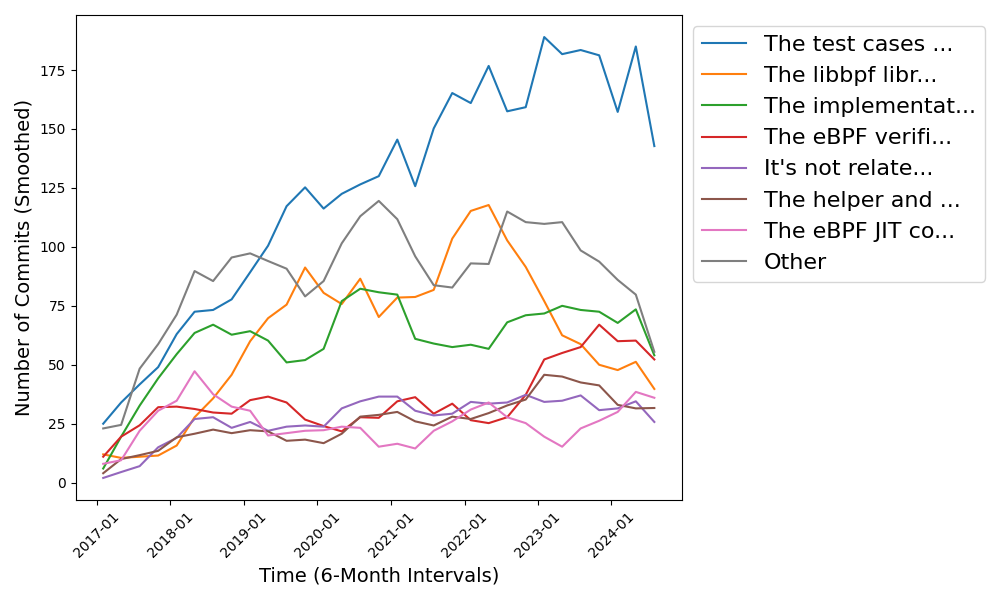}
    \caption{Commits Related to Major Implementation Components Over Time}
    \label{fig:timeline_major_related_implementation_component_smoothed}
\end{figure}

Figure~\ref{fig:timeline_major_related_implementation_component_smoothed} illustrates the evolution of major implementation components in the Linux eBPF subsystem.

Most components experienced their highest activity between 2017 and 2022, reflecting the rapid development of eBPF features during this period. The \texttt{libbpf} library saw the most dramatic increase, while the JIT compiler was most frequently updated around 2018. The rise in test cases also reflects the growing importance of a robust testing framework in this field.

After peaking around 2021--2022, several components show a decline or stabilization in activity. This indicates that many eBPF components have entered a phase of optimization and maintenance rather than new feature development, while testing continues to increase coverage. The verifier shows modest activity throughout the observed period but seems to increase in 2023--2024, which may reflect renewed research efforts in this area.

\subsubsection{Commits Related to Major Logic Components Over Time}

\begin{figure}[ht]
    \centering
    \includegraphics[width=\linewidth]{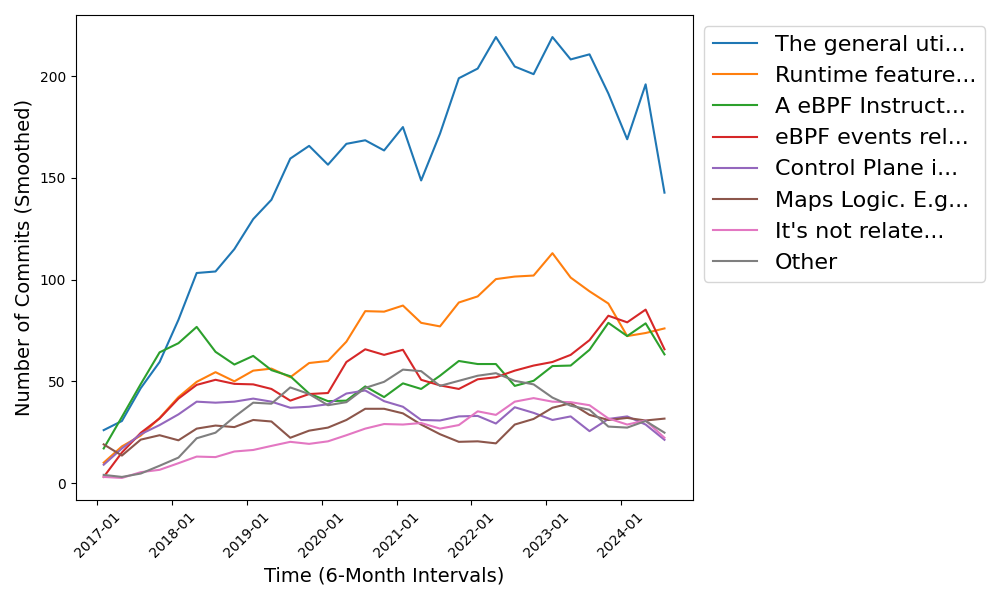}
    \caption{Commits Related to Major Logic Components Over Time}
    \label{fig:timeline_major_related_logic_component_smoothed}
\end{figure}

Figure~\ref{fig:timeline_major_related_logic_component_smoothed} illustrates the evolution of major logic components within the Linux eBPF subsystem over time. The \texttt{General Utilities} component, which includes test cases and build scripts, exhibits the most significant improvements, reaching a peak between 2022 and 2023 before experiencing a decline. In contrast, the \texttt{eBPF Instruction Logic} component displays two prominent peaks in 2018 and 2024, corresponding to the initial introduction of eBPF instructions and subsequent standardization efforts, respectively.

Other components, such as \texttt{Runtime Features, Helpers, and kfuncs}, show a notable peak in 2023 followed by a decrease and subsequent stabilization. Meanwhile, the \texttt{Control Plane Interface} and \texttt{Maps Logic} components maintain relatively steady levels of activity throughout the observed period.

\subsubsection{Use Cases or Events Over Time}

\begin{figure}[ht]
    \centering
    \includegraphics[width=\linewidth]{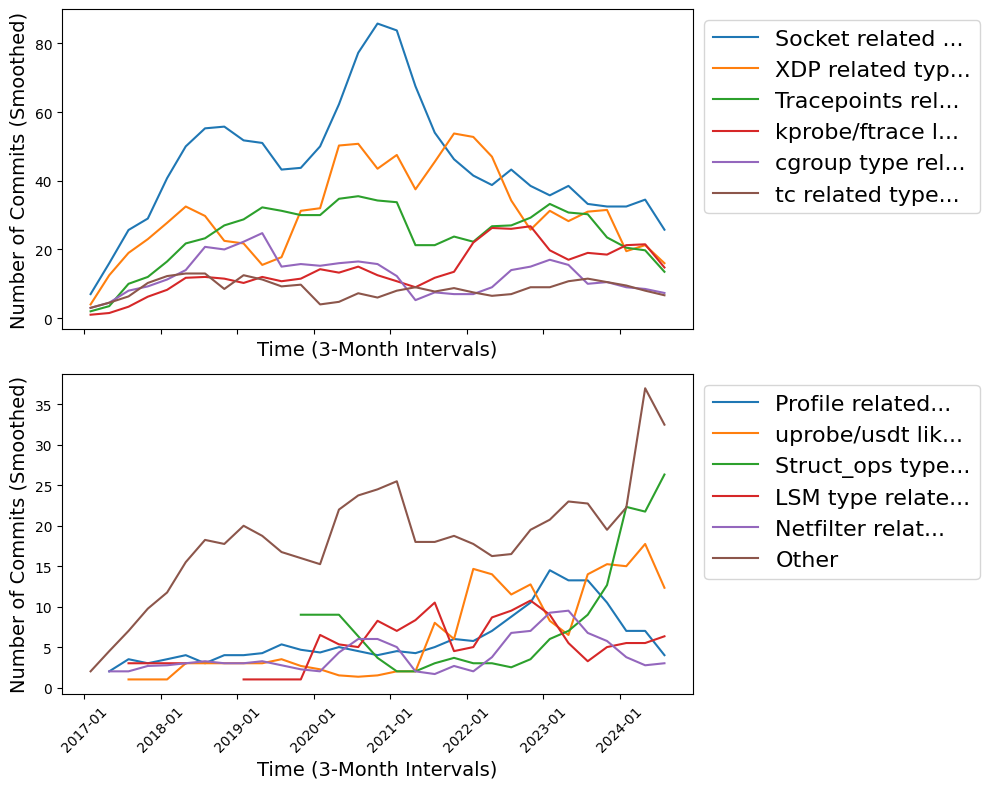}
    \caption{Use Cases or Events Over Time}
    \label{fig:timeline_usecases_or_submodule_events_smoothed}
\end{figure}

Figure~\ref{fig:timeline_usecases_or_submodule_events_smoothed} reveals significant fluctuations in event types over the years.

Notably, network-related events such as socket and XDP programs experienced a surge from 2020 to 2022, after which they entered a stabilization phase following the initial burst of feature additions and optimizations. The tracepoints-related activities show a steady increase with periodic fluctuations, peaking around 2021 before decreasing. The kprobe/ftrace-related events remain mostly stable, with a slight increase in 2022. Uprobe-related events show moderate activity over several years, with slight peaks in 2022 and again in 2024. The \texttt{struct\_ops}, an emerging feature introduced in 2020, shows a significant increase in activity between 2023 and 2024. Other events such as LSM for security remain minor.

\subsection{Deeper Insights Analysis}

This section delves into the survey responses to uncover patterns, trends, and areas for improvement within the eBPF subsystem.

\subsubsection{Which Kernel Components and Files Have the Most Frequent Bugs?}

\begin{figure}[ht]
    \centering
    \includegraphics[width=\linewidth]{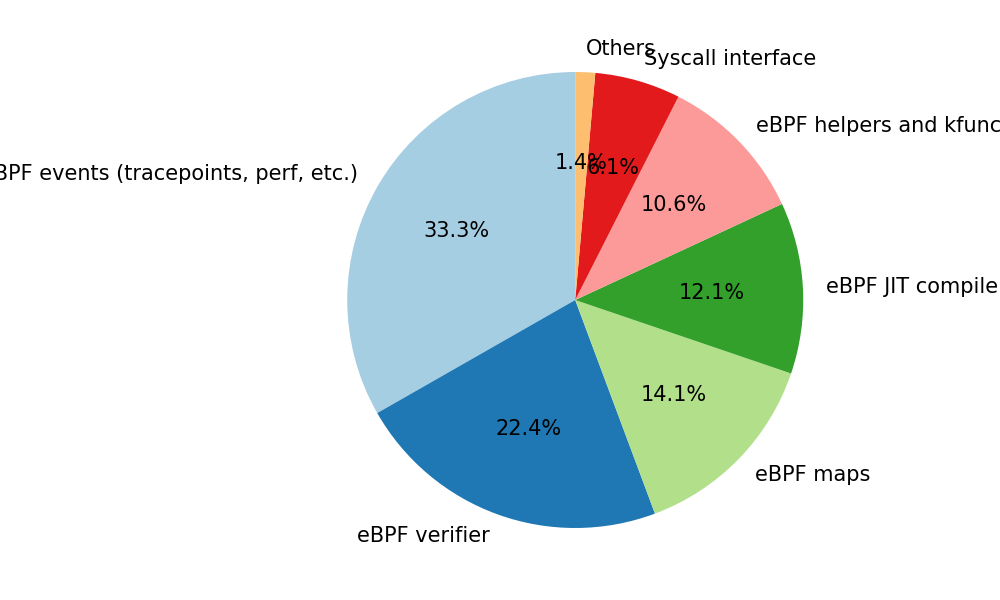}
    \caption{Kernel Implementation Components with the Most Bugs}
    \label{fig:buggy_kernel_component}
\end{figure}

Figure~\ref{fig:buggy_kernel_component} illustrates the kernel implementation components with a high number of bugs. While previous analyses and tools have primarily focused on improving the stability of the verifier and JIT compiler, these areas account for only about 35\% of the bugs. The largest number of bugs originate from eBPF event-related code, which involves the interaction of eBPF with other kernel subsystems. Additionally, helpers and maps also have a significant number of bugs. Due to the complexity of the control plane, the eBPF syscall interface is also prone to bugs.

By examining specific files, we observe that bugs frequently occur in the verifier, syscall, core, and network filter components. These files require better test coverage and more attention.

\textbf{Top 10 Buggy Files:}
\begin{verbatim}
kernel/bpf/verifier.c             425
net/core/filter.c                 140
kernel/bpf/syscall.c              111
include/linux/bpf.h                87
kernel/bpf/core.c                  83
include/uapi/linux/bpf.h           80
kernel/trace/bpf_trace.c           77
kernel/bpf/btf.c                   75
tools/include/uapi/linux/bpf.h     54
kernel/bpf/sockmap.c               51
\end{verbatim}

\subsubsection{What is the Relationship Between Instruction-Related Changes in the Verifier and All Verifier Bugs?}

\begin{figure}[ht]
    \centering
    \includegraphics[width=\linewidth]{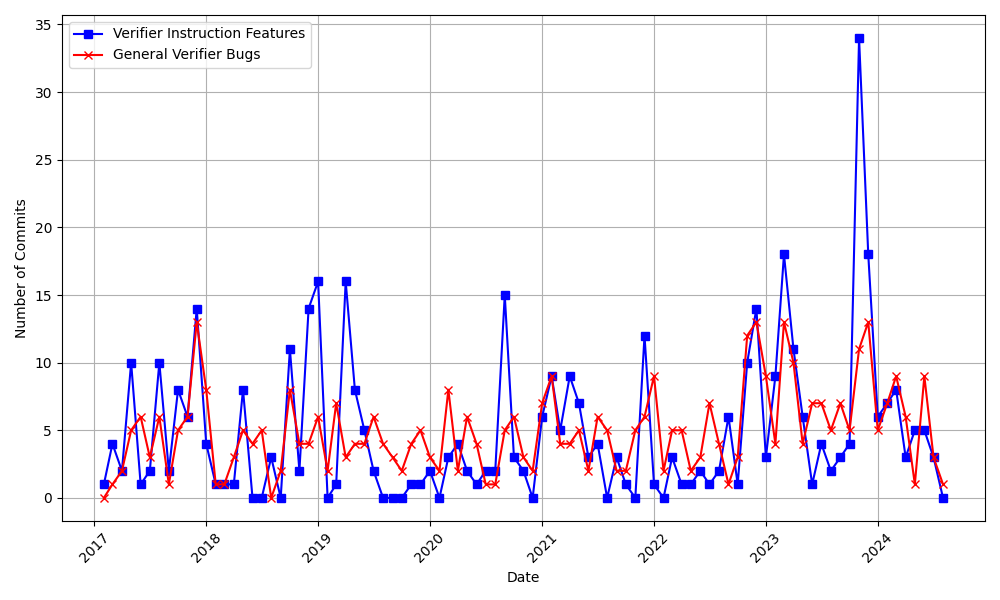}
    \caption{Verifier Bugs or Features Related to eBPF Instructions Over Time}
    \label{fig:instruction_verifier_features_bugs_over_time}
\end{figure}

Figure~\ref{fig:instruction_verifier_features_bugs_over_time} shows that changes related to eBPF instructions in the verifier closely correlate with the number of verifier bugs. This insight highlights the importance of focusing on instruction-related aspects during verifier development and debugging to enhance overall system stability.

\subsubsection{The Evolution and Status of \texttt{libbpf}}

We also examined the lifecycle of specific components, such as \texttt{libbpf}.

\begin{figure}[ht]
    \centering
    \includegraphics[width=\linewidth]{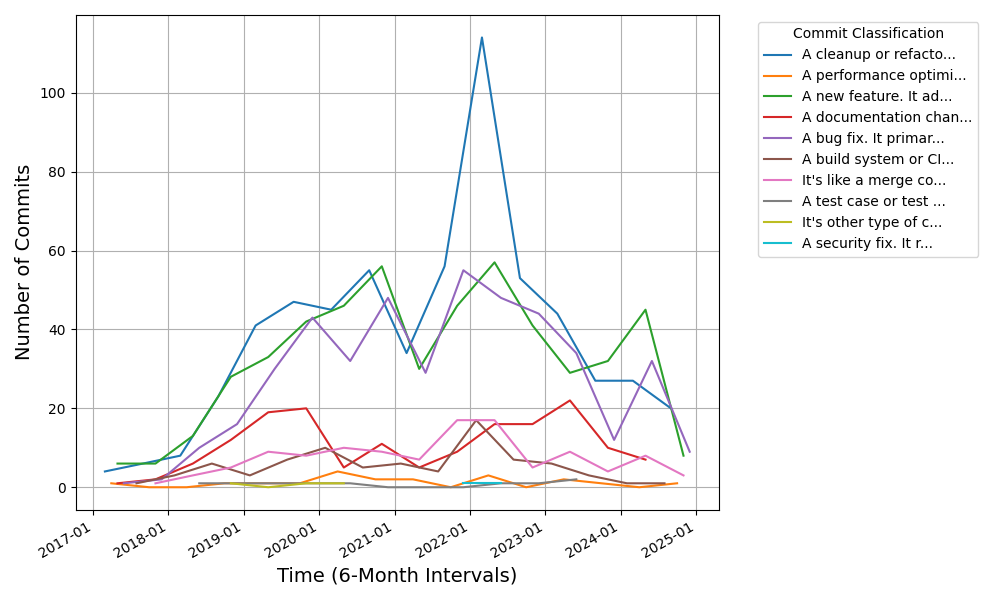}
    \caption{Evolution of \texttt{libbpf} Over Time}
    \label{fig:libbpf_commit_classification}
\end{figure}

Based on Figure~\ref{fig:libbpf_commit_classification}, the development of \texttt{libbpf} began to grow significantly in 2017. New feature development follows a cyclical pattern, with notable spikes around 2020 and 2022. After 2022, the number of cleanups and refactorings increased significantly, indicating a shift in focus towards code maintainability and stability. However, the decline in cleanup commits after 2023 suggests that while new features continue to be added, the emphasis has shifted more towards stabilization and optimization.

Historical milestones verify this trend. For instance, \texttt{libbpf} version 1.0 was released in August 2022~\cite{libbpf1}, and the major feature ``Compile Once, Run Everywhere'' (CO-RE) was introduced around 2020~\cite{core}, both aligning with the peaks and shifts observed in the commit history.

\subsubsection{What Dependencies Have Emerged Between Features and Components?}

\begin{figure}[ht]
    \centering
    \includegraphics[width=\linewidth]{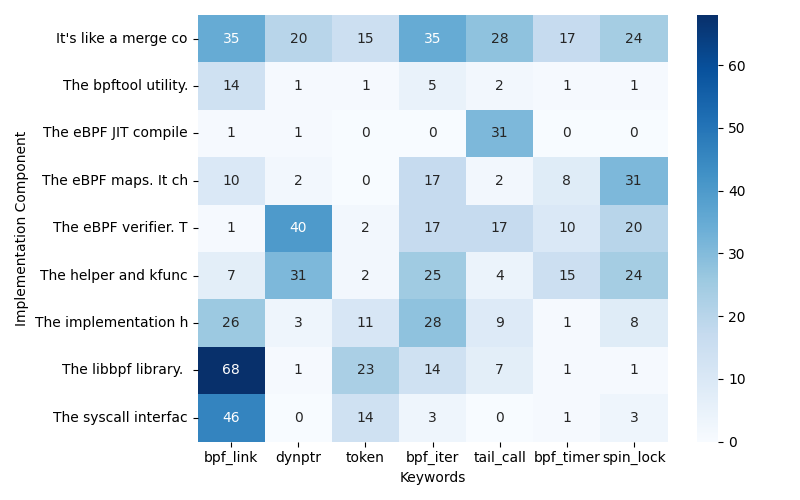}
    \caption{Feature-Component Interdependencies in the BPF Subsystem}
    \label{fig:feature_component_heatmap}
\end{figure}

Our analysis of feature-component dependencies in Figure~\ref{fig:feature_component_heatmap} reveals two primary patterns. First, new control plane abstractions such as \texttt{bpf\_link}\cite{bpflink} and \texttt{token}\cite{token} typically require coordinated updates to both the \texttt{syscall interface} and \texttt{libbpf}, indicating tightly coupled development. Second, runtime features like \texttt{bpf\_iter}\cite{bpf_iterators} and \texttt{spin\_lock}\cite{spinlock} mainly depend on internal kernel components such as \texttt{helpers} and \texttt{verifier}, with minimal impact on the \texttt{JIT compiler}, suggesting potential areas for JIT optimization.

\subsection{Expert Confirmation}

We discussed the results with more than five eBPF experts who have submitted kernel patches or presented at conferences. They confirmed that the findings align with their understanding. We also shared the report with kernel maintainers and the BPF kernel mailing list, which is currently under discussion. Additionally, we plan to provide experts with a sample of survey responses and ask them to rate the accuracy of each response on a defined scale. Based on their feedback, we will refine the survey to improve its correctness and relevance.

\subsection{Clarity and Ambiguity in Responses}

Based on the insights gathered from the survey responses, the results are generally clear and actionable. We plan to invite more experts to review a sample of the responses to identify any vague or unclear answers. Subsequently, we will revise any survey questions prone to misinterpretation or ambiguity to improve clarity and precision.

\subsection{Alignment with Real-World Changes}

By comparing survey responses with real-world feature changes discussed earlier, we confirm that the survey likely captures the correct historical context. We plan to cross-reference survey results with commit histories, mailing list discussions, and kernel release notes to further verify alignment with real-world changes.

\subsection{Coverage of Survey Questions}

We plan to review the survey's coverage to identify potential gaps, ensuring that important aspects such as security, performance, and dependencies are adequately addressed.

\section{Best Practices in the \emph{Code-Survey} Method}
\label{sec:best_practices}
The \emph{Code-Survey} methodology, particularly when integrated with Large Language Models (LLMs), requires carefully designed practices to ensure the accuracy and relevance of responses. The key principle is to create surveys intended for human participants and allow LLMs to perform on them; this approach leverages existing methodologies and can be easily adapted.

Below are some best practices to achieve reliable results:

\begin{enumerate}
    \item \textbf{Use Predefined Tags and Categories}: To minimize hallucinations or random answers from the LLM, it is essential to provide structured, closed-ended questions. Utilizing predefined tags such as ``Bug Fix,'' ``New Feature,'' or ``Performance Optimization'' helps standardize responses and reduces ambiguity.

    \item \textbf{Implement LLM Agent Workflows}: LLMs may need to review and refine their answers multiple times to improve accuracy. Incorporating feedback loops and techniques like ReAct~\cite{yao2022react} allows the model to re-evaluate its responses, enhancing overall data quality.

    \item \textbf{Allow for ``I'm Not Sure'' Responses}: Providing an option for the LLM to indicate uncertainty prevents random or misleading answers when encountering unfamiliar or complex questions. This is particularly useful in domains where the LLM's knowledge may be limited or incomplete.

    \item \textbf{Pilot Testing and Iterative Refinement}: Conducting pilot tests prior to full deployment helps identify potential issues with question clarity and LLM understanding. Iterative refinement of survey questions based on these trials ensures logical consistency and improves data reliability.

    \item \textbf{Ensure Consistency and Perform Data Validation}: Design questions for Consistency. After the survey, apply validation checks to ensure consistency across responses. Detecting and filtering out illogical or contradictory answers is essential for maintaining the integrity of the dataset.
\end{enumerate}

By adhering to these best practices, the \emph{Code-Survey} method can effectively leverage LLMs, mitigating risks such as hallucination and improving the reliability of responses in technical domains like Linux kernel commit analysis.

\section{Limitations}
\label{sec:limitations}

While \emph{Code-Survey} offers significant advancements in analyzing the evolution of large software systems, it has certain limitations:

\begin{enumerate}
    \item \textbf{Dependency on Data Quality}: The accuracy of \emph{Code-Survey} is heavily dependent on the quality and completeness of the input data. Incomplete commit messages, patches or fragmented email discussions can lead to gaps in the structured data, potentially obscuring important aspects of feature evolution.

    \item \textbf{Limitations of LLMs}: Although LLMs like GPT-4o are powerful, they are not infallible. Misinterpretations of commit messages or developer communications can result in inaccurate data structuring~\cite{ji2023survey}. LLMs may sometimes generate plausible but incorrect information (hallucinations) or miss important details in the questions~\cite{bubeck2023sparks}. To mitigate these issues, careful survey design and validation are essential to guide the model's responses more effectively.

    \item \textbf{Requirement for Human Expert Feedback}: Despite automation, human expertise remains essential for designing effective surveys, evaluating results, and ensuring the contextual relevance of the structured data. This dependency can limit the scalability of \emph{Code-Survey} in scenarios where expert availability is constrained.
\end{enumerate}

\section{Future Work}
\label{sec:future}

While \emph{Code-Survey} demonstrates significant potential in organizing and analyzing unstructured software data, several areas warrant further exploration and improvement:

\subsection{Enhanced Evaluation of LLM-Generated Survey Data}

To address challenges such as hallucinations and inaccuracies in LLM outputs, future work will focus on developing robust validation frameworks. This includes benchmarking results against curated datasets and involving human experts in refining LLM outputs. Enhancing the reliability of the structured data will improve the overall effectiveness of \emph{Code-Survey}.

\subsection{Performance Optimization with Advanced LLMs}

The current proof-of-concept showcases automation but with room for performance improvements if no human feedback is procided. Future efforts will explore the use of more advanced models, such as O1~\cite{o1}, and the implementation of multi-agent systems to optimize performance and accuracy. Ensuring compatibility with machine analysis tools is crucial for seamless integration into existing workflows.

\subsection{Application to Other Software Projects}

While \emph{Code-Survey} has been applied to the Linux eBPF subsystem, its methodology can be directly applied to other projects like Kubernetes, LLVM, and Apache. Expanding to these repositories will test its scalability and versatility, potentially requiring adjustments to accommodate different development practices and environments.

\subsection{Incorporation of Additional Data Sources like Code, Trace, and Execution Flow}

Due to the time limited, the case study currently mainly focus on analyzing the commit and features. Extending \emph{Code-Survey} to incorporate a wider range of data sources—such as source code, execution traces, and execution flows—will provide a more comprehensive understanding of software systems. Direct structuring of code and functions, transforming technical elements into structured, query-able data or graphs with attributes, will enable advanced analyses. This approach will facilitate a deeper exploration of software implementations, performance characteristics, and evolutionary patterns.

By pursuing these enhancements, \emph{Code-Survey} aims to become a comprehensive tool for analyzing complex software systems, benefiting both developers and researchers in the field of software engineering.


\section{Conclusion}

This paper introduced \emph{Code-Survey}, the first methodology that leverages LLMs for systematically exploring and analyzing large-scale codebase through a survey-based approach. Applied to the Linux eBPF subsystem, \emph{Code-Survey} successfully uncovered patterns in feature evolution and design that traditional methods overlook. Despite some limitations, our approach provides a valuable framework for understanding the growth of real-world software systems. Future work will expand its scope, enhance LLM capabilities, and apply \emph{Code-Survey} to other large-scale codebases. We also invite collaborators to work together on the ongoing development and refinement of this pioneering methodology.

\bibliographystyle{plain}
\bibliography{cite}
\end{document}